\documentclass[%
 reprint,
nofootinbib,
 amsmath,amssymb,
 aps,
]{revtex4-2}

\usepackage{graphicx}
\usepackage{amsfonts,amssymb,amstext,amsmath,amsthm,verbatim,times,cancel,epsfig,bbold}
\usepackage{hyperref}
\usepackage{xcolor}
\usepackage{enumitem}
\usepackage{dcolumn}
\usepackage{bm}

\begin{document}

\preprint{APS/123-QED}

\title{Superdilations at Schwarzschild null infinity and beyond}

\author{Marco Refuto}
\email{marco.refuto@unina.it}
\affiliation{Dipartimento di Fisica Ettore Pancini, Università di Napoli “Federico II”,
Complesso Univ.\ Monte S.\ Angelo, I-80126 Napoli, Italy}
\affiliation{INFN, Sezione di Napoli, Complesso Univ.\ Monte S.\ Angelo, I-80126 Napoli, Italy}

\begin{abstract}
Relying on the definition of asymptotic conformal Killing horizons, the generators
of asymptotic conformal symmetries at future null infinity in Schwarzschild space-time is computed.
Supertranslations and superdilations are generated by two independent functions on the sphere, and the generators close the algebra $\text{Conf}(S^2) \ltimes (C^\infty(S^2) \ltimes C^\infty(S^2))$. For a generic asymptotically flat space-time, the superdilations Iyer–Wald charge is non-trivial but divergent at the boundaries of null infinity, where its growth rate is fixed by the asymptotic values of the Bondi mass aspect. Such a rate equals the monopole supertranslation charge, for which antipodal matching is well posed and trivially satisfied in the Schwarzschild geometry, where, furthermore, only the
global dilation mode contributes. Its growth reflects the explicit time dependence of
the generator rather than radiation. The effect of superdilations on a pair of celestial observers
is an angle-dependent dilation rate, related to the local density of the charge.
\end{abstract}

\maketitle

 \section{Introduction}
Symmetries play a pivotal role in identifying connections among only apparently scattered physical phenomena, helping us highlight common features. The emergence of a symmetry structure also suggests new phenomenological paradigms to test theoretical predictions. From this perspective, the long-distance sector of classical General Relativity provides a remarkable example: the asymptotic region of several space-times of interest (such as black hole or cosmological ones) yields a non-trivial example of how results in different areas of physics are often nothing but corners of a deeper unifying structure.  Back in the sixties, Bondi, van der Burg, Metzner and Sachs (BMS) (cf. Refs \cite{Bondi:1962px, Sachs:1962zza, Sachs:1962wk}) discovered that the symmetry group of asymptotically flat space-times at null infinity is much larger than the Poincar\'e one, having a group structure
\begin{equation}
    \text{BMS}_4=\text{SO}^+(3,1)\ltimes C^\infty(S^2),
\end{equation}
where the $\text{SO}^+(3,1)$ part encodes Lorentz transformations\footnote{We do not consider possible extensions of the BMS group. For a review, cf. Refs \cite{Compere:2019qed, Donnay:2023mrd}.}, while the Abelian Poincar\'e subgroup is promoted to angle-dependent translations, called supertranslations. We have therefore that asymptotic symmetries are not only non-trivial, but, as later discovered, deeply related to observable effects. In the same years Weinberg proved the soft graviton theorem in Ref. \cite{Weinberg:1965nx}: the addition of a low energy (soft) graviton to an external leg of any Feynman diagram is equivalent to multiply it by an overall soft factor. Later, in the seventies, Braginsky and Thorne in Ref. \cite{Braginsky:1987kwo} described the gravitational wave displacement memory effect: two detectors on the same celestial sphere are permanently shifted after the passage of a gravitational or radiation wave. These three results share only one property: they describe the low energy regime, also called infrared, of gravity. Strominger et al. (cf. Refs \cite{Strominger:2013jfa, Strominger:2014pwa, He:2014laa}), over a decade ago,  proved that such outcomes are corners of what is called "infrared triangle": a connection between apparently disconnected theoretical frameworks which becomes manifest in the asymptotic sector of gravity. We have therefore that, after several decades from the first studies of the asymptotic structure of space-times, this topic has re-emerged as a fertile arena to test new physics thanks to the relations provided by the infrared triangle. The structure described above concerns supertranslations: several other instances have been reported in the literature for Abelian or non-Abelian gauge theories, as described in Ref. \cite{Strominger:2017zoo}. 
Thus far, we have seen that the asymptotic symmetry group of asymptotically flat space-times is an enhanced version of the Poincar\'e group. Following this pattern, it appears quite natural to ask whether it is possible to enlarge dilations too. This happens for de Sitter-like space-times, as discovered by Dappiaggi, Moretti and Pinamonti (cf. Ref. \cite{Dappiaggi:2017kka}). In searching for the counterpart of the BMS group for expanding FRW space-times, they introduce the definition of expanding universe with geodesically complete cosmological particle horizon, which offers a generalisation of the usual homogeneous or isotropic space-times used to model an expanding universe. In this class of space-times, they found that the horizon enjoys a BMS - like symmetry group, called the DMP group. It has the following structure
\begin{equation}
    \text{DMP}_4=\text{SO}(3) \ltimes (C^\infty(S^2)\ltimes C^\infty(S^2)),
\end{equation}
where SO$(3)$ corresponds to the isometry group of the sphere\footnote{The DMP group is not a generalisation of the full de Sitter group $\text{SO}^+(4,1)$, since it does not preserve the horizon, but instead of the cosmological de Sitter group. It contains all the generators preserving the cosmological horizon, ruling out de Sitter boost generators}, and the two $C^\infty(S^2)$ factors encode supertranslations and superdilations. The horizon Lie algebra is the infinite-dimensional Lie algebra of smooth vector fields on $\mathbb{R}\times S^2$ generated by, besides the Killing vector fields on the sphere, 
\begin{equation}
    \beta \partial_u, \quad \alpha u\partial_u 
\end{equation}
with $\alpha,\beta \in C^\infty(S^2)$ and $u$ is a time coordinate. However, superdilations do not arise only in such a scenario. In Ref. \cite{Donnay:2020fof}, starting from the Minkowski metric in Bondi coordinates but with a compactified radial coordinate $\rho=1/r$ in $d$ dimensions
\begin{equation}
    ds^2=-du^2+\frac{1}{\rho^2}(2dud\rho+d^{d-2}\Omega^2),
\end{equation}
the authors computed the Killing and conformal Killing vector field in the limit $\rho=0$, defining them asymptotic. They obtain, besides supertranslations and superrotations (cf. Eqs (2.23) in Ref. \cite{Donnay:2020fof}), an asymptotic superdilations vector field 
\begin{equation}
    \xi_D(g)|_{\rho=0}=g u\partial_u,
\end{equation}
with $g\in C^\infty(S^2)$. Moreover, they extend the analysis to Minkowski conformally related metrics (such as $\text{AdS}_d$, $\text{dS}_d$), mapping Minkowski flat null infinity $\mathcal{I}^+$ into horizons $H_+$ of conformally flat space-times. These examples exhibit the emergence of superdilations in Petrov O metrics and their generalisations. It would be of interest to study if superdilations exist in the asymptotically flat case, as it may enlarge our understanding of the infrared sector of gravity in black hole space-times. The general interest in such a research area is mainly related, e.g., to the study of infrared triangles and the relation between asymptotic symmetries and black hole soft hair (cf. Ref. \cite{Hawking:2016msc}). In this work, we aim at explicitly computing the algebra of asymptotic conformal symmetries at Schwarzschild null infinity by exploiting the geometrical construction of asymptotic conformal Killing horizons given in Ref. \cite{deAguiarAlves:2025dce}.  The algebra we obtain extends the BMS one, resembling the DMP algebra (as theoretically described in Ref. \cite{deAguiarAlves:2025dce}). 
The asymptotic conformal group is given by Eq. \eqref{asymptotic conformal group Scwharzschild} and it aims at offering an asymptotic generalisation of conformal symmetries. 
However, these results are beyond a pure mathematical interest. 
As already stated, the physics underlying supertranslations and superrotations in asymptotically flat space-times concerns memory effects. The effect of superdilations on a pair of celestial observers is considered in Sec. VII:  their displacement is governed by an angle-dependent dilation rate. Similar results can be found in extended theories of gravity\footnote{In General Relativity, gravitational waves have transverse and traceless polarisations: the two characteristic modes of expansion for a congruence of geodesics are the plus ($+$) and cross ($\times$) mode.}.  Notably, in Brans-Dicke gravity (cf. Ref. \cite{Brans:1961sx}) with a massless scalar field, one possible gravitational-wave polarisation is a transverse excitation which expands and contracts an array of inertial particles. Furthermore, the theory predicts a memory effect beyond General Relativity and it is possible to provide a connection with soft theorems and asymptotic symmetries (cf. Refs \cite{Hou:2020tnd, Seraj:2021qja, Campiglia:2017dpg, Campiglia:2018see}).  In our analysis we obtain a very similar effect on an ensemble of BMS observers (throughout the angle-dependent dilation rate), without adding fields propagating in the space-time, but instead considering only its asymptotic symmetrical structure, i.e., its geometric invariance properties. Symmetries at infinity therefore can play a relevant role in identifying which degrees of freedom are fundamental in a theory of gravity. 
In this paper, we rely on the abstract indices notation (cf. Ref. \cite{Wald:1984rg}) and, if not specified, we work in natural units $(c=G=1$). We consider four-dimensional space-times with metric signature $(-,+,+,+)$. Capital Latin letters label the standard angular variables, which we consider ranging in the interval\footnote{The choice of the semi-open interval $[0, 2\pi)$ for the azimuthal variable $\phi$ avoids overlapping boundaries, thereby satisfying the requirements for a well-defined coordinate chart across the continuous manifold.} 
\begin{equation}
    \{\theta,\phi\}\in[0,\pi]\times [0,2\pi).
    \label{range of angular variables}
\end{equation}
$D_A$ labels the covariant derivative on the two sphere. The paper is organised as follows.\\
In Sec. II we survey the definition of asymptotically flat space-times and the derivation of the BMS algebra.\\
In Sec. III we discuss a very standard but often neglected result in General Relativity: asymptotically flat space-times, in most cases, do not admit conformal Killing vector fields. Since dilations belong to the set of conformal transformations, this is the first obstacle in looking for their angle-dependent generalisation: their existence is highly non-trivial. \\
In Sec. IV we provide an overview of the definitions introduced in Ref. \cite{deAguiarAlves:2025dce} in order to describe the underlying geometric framework we adopt in this paper. The main computation concerns solving the conformal Killing equation near null infinity, i.e., asymptotically. Our derivation is general. The restriction to constant Bondi mass aspect is needed only for the explicit subleading analysis. \\
In Sec. V we derive the generator of asymptotic conformal symmetries (cf. Eq. \eqref{final xi vector}), and its algebra. 
In Sec. VI we compute the Iyer-Wald charge, first for a generic asymptotically flat space-time and then for Schwarzschild.
\\
In Sec. VII we study the effect of superdilations on a pair of BMS observers. This new symmetry induces what we have identified as an angle-dependent dilation rate. A relation with the Iyer-Wald charge is then considered.\\

\section{Asymptotic flatness and symmetries at null infinity}

Throughout all physically relevant space-times, asymptotically flat ones underpin several fundamental aspects of physics, as they offer the gravitational equivalent of an isolated system. The most important example is given by black hole metrics: they constitute an excellent arena for probing theoretical models and potential experimental scenarios. The underlying idea of an asymptotically flat space-time is that, far away from a matter source, it resembles Minkowski. However, there are five different kinds of infinity in Minkowski space-time (past and future timelike and null infinity, spacelike infinity). We are interested in studying null infinity: the region reached by radiation and gravitational waves (or, from a particle perspective, by photons and gravitons). There are two equivalent ways to mathematically define asymptotic flatness. A covariant geometrical description was first provided by Penrose in Ref. \cite{Penrose1963}. A four-dimensional Lorentzian space-time $(M,g_{ab})$ is said to be asymptotically flat at future null infinity with conformal extension $(\tilde M, \tilde g_{ab})$ and conformal factor $\Omega$ if, given the smooth embedding $\psi: M\rightarrow \tilde M$ with open range and a smooth function $\Omega: \psi(M)\rightarrow \mathbb{R}$ such that $\Omega>0$ and its pullback satisfies
\begin{equation}
    \psi^* \tilde g_{\psi(M)}=\Omega^2 g_M,
\end{equation}
certain requirements concerning regularity and causality are satisfied 
(see, e.g., Ref. \cite{Wald:1984rg}). They govern the relationship between the bulk $(M,g_{ab})$ space-time and its unphysical conformal extension $(\tilde M, \tilde g_{ab})$. Notice that the definition is mainly based on the conformal transformation governed by $\Omega$. This scheme offers a powerful framework for studying global properties, especially existence theorems. Another route is provided by a coordinate-dependent approach: we impose a priori a gauge, manifestly breaking the diffeomorphism invariance of the theory, in order to introduce a "good" coordinate system to study null infinity. The Bondi gauge \cite{Compere:2019qed} generalises to arbitrary space-times null-spherical Minkowski coordinates, used for instance to define the celestial sphere. The existence of such a coordinate system for a given space-time ensures that it is asymptotically flat. Despite the loss of diffeomorphism invariance, such a procedure is better suited to explicit computations and to the study of local properties, such as the algebra, charges, and observable effects of asymptotic symmetries. Since we aim at studying this kind of features, we rely on this Bondi-gauge description. We therefore introduce retarded\footnote{At past null infinity, $\mathcal{I^-}$, the advanced time coordinate $v$ is considered instead of the retarded time $u$. } Bondi coordinates $(u,r,x^A)$, ranging in the interval
\begin{equation}
    \{u,r\}\in (-\infty,\infty)\times [r_-,r_+]
\end{equation}
(where $r_+$ can be infinite),  while the angular coordinates $x^A=\{\theta,\phi\}$ range in the interval described in Eq. \eqref{range of angular variables}.
The metric of asymptotically flat space-times, near future null infinity $\mathcal{I}^+$,  is written as (cf., e.g., Refs \cite{Strominger:2017zoo, Compere:2019qed, Kervyn:2023adk})
\begin{align}
ds^2=&-\left(1-\frac{2m}{r}\right)du^2-\left(2-\frac{C_{AB}C^{AB}}{16r^2}\right)dudr \nonumber \\&+D^BC_{AB}dudx^A \nonumber\\&+\frac{1}{r}\left[\frac{4}{3}(N_A+u\partial_Am)-\frac{1}{8}\partial_A(C^{BC}C_{BC})\right]dudx^A\nonumber\\&+\left(r^2\gamma_{AB}+rC_{AB}+\frac{1}{4}\gamma_{AB}C_{CD}C^{CD}\right)dx^Adx^B\nonumber\\
&+ \text{subleading terms},
\label{metric in Bondi gauge}
\end{align}
where $m(u,x^A)$ is the Bondi mass aspect, $C_{AB}(u,x^A)$ is a traceless and symmetric field encoding information on radiation at $\mathcal{I}^+$ and $N_A(u,x^A)$ is the angular momentum aspect. 
These functions are constrained by Einstein equations. In vacuum, one of these relations is the mass-loss equation, given by
\begin{equation}
    \partial_um=\frac{1}{4}D_AD_BN^{AB}-\frac{1}{8}N_{AB}N^{AB}
\end{equation}
As already pointed out in the introduction, the symmetry group at null infinity is much larger with respect to the Minkowski one. In order to obtain its generators, there must exist a vector field $\xi^a$ which preserves the Bondi gauge, i.e., 
\begin{align}
    &(\mathcal{L}_\xi g)_{rr}=0, \quad  
    (\mathcal{L}_\xi g)_{rA}=0, \nonumber\\   
    &\mathcal{L}_\xi \left[\partial_r\text{det}\left(\frac{g_{AB}}{r^2}\right)\right]=0 \label{Bondi gauge conditions},
\end{align}
within the fall-off conditions
\begin{align}
     &(\mathcal{L}_\xi g)_{uu}=O(r^{-1}),  \quad  (\mathcal{L}_\xi g)_{ur}= O(r^{-2}),\nonumber\\  &(\mathcal{L}_\xi g)_{uA}=O(r^{0}), \quad (\mathcal{L}_\xi g)_{AB}=O(r) \label{Bondi gauge condition falloff}.
\end{align}
Such a vector field is (neglecting subleading terms)
\begin{align}
    \xi=&\left[ g+\frac{u}{2}D_AY^A  \right]\partial_u \nonumber\\
    &+\left[Y^A-\frac{1}{r}D^Ag-\frac{u}{2r}D^AD_BY^B\right]\partial_A  \nonumber\\
    &+\left[-\frac{r+u}{2}D_AY^A+\frac{1}{2}D_AD^Ag\right]\partial_r.
    \label{BMS generator xi}
\end{align}
$g=g(x^A)$ is an unconstrained function on the two-sphere, generating supertranslations (angle-dependent translations). $Y^A=Y^A(x^B)$ is a conformal vector field on the two sphere, generating the Lorentz sector. Since the canonical charges associated with these generators are non-trivial, these diffeomorphisms acquire the name of asymptotic symmetries. 
They can be regarded as \cite{Strominger:2017zoo}
\begin{equation}
    \text{Asymptotic symmetries} = \frac{\text{allowed gauge symmetries}}{\text{trivial gauge symmetries}}.
\end{equation}
Upon splitting the supertranslation (denoted by a sub-index $_T$) sector from the Lorentz ($_R$) one
\begin{align}
    \xi_T(g)=&g\partial_u-\frac{1}{r}D^Ag\partial_A\nonumber\\&+\frac{1}{2}D_AD^Ag\partial_r,\\
    \xi_R(Y)=&\frac{u}{2}D_AY^A\partial_u+Y^A\partial_A\nonumber\\&-\frac{r+u}{2}D_AY^A\partial_r,
\end{align}
it is easy to compute the BMS algebra, given by
\begin{align}
    [\xi_T(g_1), \xi_T(g_2)] &= 0, \label{BMS algebra1} \\
    [\xi_R(Y_1), \xi_R(Y_2)] &= \xi_R([Y_1, Y_2]), \label{BMS algebra2} \\
    [\xi_R(Y), \xi_T(g)]   &= \xi_T\left(Y[g]-\frac{1}{2}gD_AY^A\right), \label{BMS algebra3}
\end{align}
where  $[Y_1,Y_2]^A=(\mathcal{L}_{Y_1}Y_2)^A=Y_1^B\partial_B Y^A_2-Y^B_2\partial_BY^A_1$ and  $Y[g]=\mathcal{L}_Yg=Y^A\partial_Ag$.  
In the following, we denote the vector field in Eq. \eqref{BMS generator xi} as $\xi_{\text{BMS}}^a$. 

\section{No conformal Killing vector fields in asymptotically flat space-times}
We briefly review some relevant results on conformal symmetries in asymptotically flat space-times. 
The motivation stems from a comparison with supertranslations: these constitute an asymptotic, angle-dependent generalisation of a symmetry already present in the bulk of the space-time. Black-hole metrics, on the other hand, do not admit dilations as bulk symmetries. Superdilations, therefore, if they exist, would not share such a connection with the interior region of the space-time. Given a space-time metric $g_{ab}$ and a vector field $X^a$ satisfying the equation
\begin{equation}
    (\mathcal{L}_X g)_{ab}=\Lambda g_{ab},
\end{equation}
$X^a$ is a conformal, homothetic or proper Killing vector field if $\Lambda$ is a function, a constant or it vanishes, respectively. In the case of a conformal Killing vector field, which we label as proper conformal Killing vector field, $\Lambda=\frac{1}{2}\nabla_aX^a$. 
The existence of a proper conformal Killing vector field  is a demanding requirement for the space-time structure and it is guaranteed only in a few cases.  Garfinkle \cite{garfinkle1987}   analysed the case of a space-time which is asymptotically flat at null infinity. If it is vacuum, asymptotically Minkowskian and it has positive Bondi energy (see the reference for the appropriate definition) for every cross section of $\mathcal{I}^+$, then it does not admit a conformal Killing field that is not a proper Killing field. Even if such a result is solely based on the structure at null infinity and on a condition on the Bondi energy, it can be extended further. By using the positive energy condition and the $3+1$ formalism,  Eardley et al.  \cite{Eardley:1986en} showed that the only solution of the Einstein vacuum equation, asymptotically flat (spatially) and admitting a conformal Killing field, is the Minkowski space-time. Moreover, they proved that the only solution with a proper homothetic symmetry reads as
\begin{equation}
    ds^2=e^{\alpha t}(-dt^2+h_{ab}dx^adx^b),
    \label{Eardley paper first metric}
\end{equation}
where $h_{ab}dx^adx^b$ is a $3$-dimensional Riemannian metric of constant negative curvature on a compact manifold, while $\alpha$ is a constant. This is the case of an expanding hyperbolic space-time. In addition (cf. Theorem $3$ in Ref. \cite{Eardley:1986en}), all the space-times with proper conformal symmetries are either everywhere locally flat (like the previous one) or else are of a particular algebraically special form,
\begin{equation}
    ds^2=-2H(u,x^C)du^2-2dudr+\delta_{AB}dx^Adx^B.
    \label{Eardley paper second metric}
\end{equation}
The vacuum Einstein equations require that the function $H$ satisfies the $2$-dimensional Laplace equation in the $\{x^C\}$, i.e., $\delta^{AB}\partial_A \partial_B H(u,x^C)=0$. These are plane-fronted waves.
Di Prisco et al. \cite{DiPrisco1987}, relied on a coordinate approach to study the role of conformal symmetries in vacuum Bondi-type metrics. Excluding all the singular solutions that do exist, the only completely regular (on the sphere) metric, compatible with the existence of a conformal Killing vector field, is the Minkowski metric.  It turns out that the existence of a one-parameter group of conformal motions appears to be a too restrictive condition if we consider the family of physically significant Bondi metrics. Indeed, in the paper they have used the leading terms in the expansion of the Bondi metric to show that the existence of a conformal Killing vector field implies that the space-time is either flat (Minkowski) or non-globally regular at null infinity. Once again, it is important to stress that solutions presenting angular singularities (unbounded sources) and/or metrics corresponding to systems radiating during an infinite period of time do, in principle, exist. All of these scattered results can be easily grouped by considering the Petrov classification (cf., e.g., Chap. IV of Ref. \cite{Stephani:2003tm}). Sharma, in Ref. \cite{sharma_1988}, studied solutions which admit proper conformal symmetries. Unlike isometries and homothetic symmetries, proper conformal symmetries do not preserve the Einstein tensor. The existence of a proper conformal symmetry therefore restricts the conformal symmetric space-times (which are, in general, of type Petrov O, N or D) to be type Petrov O (Minkowski or conformally related) or N (plane-fronted waves) only. 
Finally, in a more recent paper, Keane \cite{Keane:2020vat} proved by a direct computation that
any conformal Killing vector field admitted by the Kerr-Newman space-time is necessarily a proper Killing vector field. This means that the most studied black-hole metrics (Schwarzschild, Reissner-Nordstroem, Kerr and Kerr-Newman) admit only the Lie algebras of proper Killing vector fields as their maximal point symmetry algebras.

\section{Asymptotic conformal Killing horizons}
The definition of asymptotic conformal Killing horizons (ACKHs) is given in Ref. \cite{deAguiarAlves:2025dce}. Such horizons admit the existence of conformal symmetries in asymptotically flat space-times, but only in the asymptotic region. In order to introduce the geometrical emergence of superdilations, we give a brief review of the main definitions.

\subsection{Basic definitions}
Suppose that, given a $d$-dimensional Lorentzian space-time (modulo diffeomorphisms invariance) $(M,g_{ab})$, it exists a manifold $\tilde M$ and an embedding $\psi: M\rightarrow \tilde M$. A null surface $\mathcal{N}$ contained in the closure of $M$ and $\tilde M $ is an ACKH for $(M,g_{ab})$ if there exists a vector field $X^a$, called asymptotic conformal Killing vector field, with the following properties:
\begin{enumerate}[label=\roman*.]
    \item $X^a$ can be extended to $\mathcal{N}$ with complete integral lines thereon;
    \item $X^aX_a$ can be extended to $\mathcal{N}$ and vanishes thereon;
    \item   There exists a function $\Omega\in C^\infty(M)$ satisfying \begin{equation}
        \mathcal{L}_X(\Omega^2)=-\frac{2}{d}\Omega^2\nabla_aX^a
    \end{equation} and such that $(\mathcal{L}_x\Omega^2g)_{ab}$ can be extended to $\mathcal{N}$ vanishing thereon. 
\end{enumerate}
According to this definition\footnote{It is interesting to stress that, following this scheme, the conformal structure at infinity is not dictated by the conformal factor $\Omega$, as it happens in the canonical construction (cf. Ref. \cite{Wald:1984rg}). Superdilations come from this higher degree of freedom.}, the existence of an ACKH relies mainly on the existence of a vector field which solves the conformal Killing equation only when restricted to $\mathcal{N}$, i.e.,
\begin{equation}
(\mathcal{L}_Xg)_{ab}|_{\mathcal{N}}=\Lambda g_{ab},
\end{equation}
where $\Lambda=\frac{1}{2}\nabla_aX^a$. The requirement of a null $X^a$ (cf. Axiom (ii)) comes directly from the classical definition of a Killing horizon: a surface where the Killing vector field becomes null and it is orthogonal to such a surface. The existence of complete integral lines assures that the ACKH has a regular global topology, ruling out geometrical singularities.
From now on we focus our attention on the $\mathcal{N}=\mathcal{I}^+$ case, since null infinity is itself an ACKH.
The horizon is however not unique. Indeed, there are several choices of  $\Omega$ such that, for $(\tilde M, \Omega^2 g_{ab})$, $\mathcal{N}$ is an ACKH. In order to take into account the existence of such a  conformal freedom, we consider null infinity as the first element of a triplet named Carrollian structure (cf. Ref. \cite{deAguiarAlves:2025dce} and references therein) and given by 
\begin{equation}
    (\mathcal{N}, \tilde h_{ab},\tilde n^a),
\end{equation}
where $\tilde h_{ab}$ is the metric induced by $\tilde g_{ab}=\Omega^2 g_{ab}$ on $\mathcal{N}$ and $\tilde n^a$ is a geodesic generator obtained as follows. Let us consider the covariant derivative $\tilde \nabla_a$ associated to the compactified metric $\tilde g_{ab}$. The inaffinity $k$ for the asymptotic conformal Killing vector field $X^a$ is defined as
\begin{equation}
    X^b\tilde \nabla_bX^a=kX^a.
\end{equation}
Its physical interpretation is given by noting that $k$ is the surface gravity \cite{Wald:1984rg}.
By introducing the $\sigma$ function, defined as
\begin{equation}
    \mathcal{L}_X\sigma=k,
\end{equation}
it is possible to compute the vector field
\begin{equation}
    \tilde n^a=e^{-\sigma} X^a
    \label{equation defining n tilde}
\end{equation}
such that
\begin{equation}
    \tilde n^b\tilde\nabla_b \tilde n^a=0,
    \label{defining property of tilde n}
\end{equation}
 i.e., $\tilde n^a$ is the affinely parametrized tangent to the null geodesic generators of the ACKH. Since the Carrollian structure preserves its properties under every conformal rescaling $\Omega$ such that $\mathcal{L}_X\Omega=0$, we have an equivalence relation among conformally related Carrollian structures
\begin{equation}
    (\mathcal{N}, \tilde h_{ab},\tilde n^a)\sim (\mathcal{N}, \omega^2\tilde h_{ab},\sigma\tilde n^a),
    \label{equivalence class of Carrollian structures}
\end{equation}
for $\omega,\sigma \in C^\infty(\Sigma)$, where $\Sigma$ is the cross section of $\mathcal{N}$. The canonical compactification leads to $\sigma=\omega^{-1}$ (or, equivalently, to use the conformal factor to define $\tilde n^a$ such as $\tilde n^a=\tilde g^{ab}\tilde \nabla_b \Omega$). Such a choice rules out superdilations a priori. The core of this geometrical picture is the freedom in the choice of  $\omega$ and $\sigma$, which are two independent functions. From the above equivalence relation we are able to derive the generator of asymptotic conformal symmetries. Indeed, such a generator is provided by every vector field $\xi^a$ satisfying the equations (infinitesimal form of the equivalence relation in Eq. \eqref{equivalence class of Carrollian structures})
\begin{align}
    (\mathcal{L}_\xi\tilde h)_{ab}=&\mu \tilde h_{ab}, \label{lee derivative of the metric at null infinity wrt xi}\\
    (\mathcal{L}_\xi\tilde n)^a=&\lambda \tilde n^a, \label{lee derivative of tilde n wrt xi definition}
\end{align}
since it acts as a conformal transformation on the induced metric $\tilde h_{ab}$ and it preserves $\tilde n^a$ along its flux up to a scaling governed by the function $\lambda$.
We stress that $\mu$ and $\lambda$ are functions on $\mathcal{I}^+$ constant along the generators. An argument supporting such a statement is the following. Since $\tilde n$ is affinely parametrised, this property still holds after the rescaling
\begin{equation}
    (\lambda \tilde n)\cdot \tilde \nabla(\lambda \tilde n)=\lambda \tilde n(\lambda) \tilde n
    \label{rescaling tilde n}
\end{equation}
if and only if $\tilde n(\lambda)=0$. Without such a condition (we provide a proof in Schwarzschild space-time in the next section), $\xi^u(u,x^A)$ would be an arbitrary function of $u$, leading to an infinite-dimensional reparametrisation algebra along each generator, and not an extension of BMS. The associated symmetry group generated by $\xi^a$ has the structure
\begin{equation}
    G_{\text{ACKH}}
=
\text{Conf}(\Sigma)
\ltimes
\big( C^\infty(\Sigma) \ltimes C^\infty(\Sigma) \big),
\label{symmetry group for ACKH}
\end{equation}
where Conf$(\Sigma)$ is the conformal group of $(\Sigma, \tilde h_{ab})$. We obtain therefore an extended version of the BMS group by including new transformations in the form of superdilations\footnote{We have in fact two $C^\infty(\Sigma)$ pieces, one for supertranslations and the other for superdilations.}. It is worth highlighting the main difference among $X^a$ and $\xi^a$, since these two vector fields are of main interest in the following. The existence of the asymptotic conformal Killing vector field $X^a$ guarantees, by definition, the existence of a ACKH $\mathcal{N}$. However, such a horizon is only one of the three elements which completely define the geometrical Carrollian structure of asymptotically flat space-times at null infinity (cf. Eq. \eqref{equivalence class of Carrollian structures}). The vector field $\xi^a$  is then the generator of asymptotic conformal transformations of the full structure at null infinity where, besides $\mathcal{N}$, it also includes an induced metric $\tilde h_{ab}$ and the vector field $\tilde n^a$. We therefore have to compute, as first step, $X^a$. The final algebra of the asymptotic symmetry group is then fully encoded in $\xi^a$ (which depends indirectly on $X^a$).

\subsection{The asymptotic conformal Killing equation }

In order to derive explicitly the asymptotic conformal Killing vector field $X^a$, we have to compute the generic vector field
\begin{align}
X(u,r,x^A)=&X^u(u,r,x^A)\partial_u+X^r(u,r,x^A)\partial_r\nonumber\\&+X^A(u,r,x^B)\partial_A
\end{align}
satisfying the conformal Killing equation at null infinity 
\begin{equation}
    (\mathcal{L}_Xg)_{ab}|_{\mathcal{I}^+}=\Lambda g_{ab},
    \label{conformal Killing equation for X}
\end{equation}
where $g_{ab}$ is the metric written in Bondi gauge as given in Eq. \eqref{metric in Bondi gauge}. Since this equation must be satisfied at null infinity, we aim at solving it by considering, for every component, the leading and first subleading\footnote{By considering only the leading term, one would simply obtain $X^a=\xi^a_{BMS}$. We have therefore to consider at least the next subleading correction.} order in $r$. The first constraints on $X^a$ can be easily read from the $rr$, $rA$ and $uA$ components of Eq. \eqref{conformal Killing equation for X}:
\begin{align}
    (\mathcal{L}_Xg)_{rr}=0&\implies \partial_rX^u=0,\\
    (\mathcal{L}_Xg)_{rA}=0&\implies  \partial_rX^A=\frac{1}{r^2}D^AX^u,\\
    (\mathcal{L}_Xg)_{uA}=\Lambda g_{uA}&\implies \partial_uX^A=0,
\end{align}
leading to
\begin{align}
X(u,r,x^A)=&X^u(u,x^A)\partial_u+X^r(u,r,x^A)\partial_r\nonumber\\&+\left[Y^A(x^B)-\frac{1}{r}D^AX^u\right]\partial_A.
    \label{general form of X after the first three constraints}
\end{align}
The $uu$, $ur$ and $AB$ components of Eq. \eqref{conformal Killing equation for X} leads to the PDEs, respectively,
\begin{align}
    2\partial_uX^u+2\left(1+\frac{2m}{r}\right)\partial_uX^r &+\frac{2m}{r^2}X^r\nonumber\\-\frac{2}{r}(X^u\partial_um+X^A\partial_Am)&=\Lambda+O(r^{-2}), \label{first PDE for X}\\
    \partial_uX^u+\partial_rX^r&=\Lambda,\label{second PDE for X}\\
    (\mathcal{L}_X\gamma)_{AB}&=\left(\Lambda-\frac{2}{r}X^r\right)\gamma_{AB}+\frac{1}{r}\Gamma_{AB},\label{third PDE for X}
\end{align}
where, in Eq. \eqref{third PDE for X} , we have highlighted the subleading correction term
\begin{equation}
    {\Gamma}_{AB}=\Lambda C_{AB}-(\mathcal{L}_XC)_{AB},
\end{equation} 
The simplest class of space-times in which these equations are non-trivially solvable  is such that
\begin{equation}
    m(u,x^A)=M, \quad \Gamma_{AB}=0,
    \label{conditions in which we solve asymptotic conformal killing equation}
\end{equation}
i.e., the Bondi mass aspect is constant and the shear tensor $C_{AB}$ is constrained through the equation $\Gamma_{AB}=0$.  There are several ways to satisfy this last requirement. The easiest one is to naively impose $C_{AB}=0$.
If we further restrict to this family of space-times, we are considering non-radiating, stationary vacuum solutions where the trivial one is provided by Minkowski space-time. The others are, e.g., black hole Petrov D metrics, such as the Schwarzschild or Kerr (and their charged version) metric. We therefore consider Schwarzschild space-time as the representative case study. In addition, the asymptotic structure does not depend on all the bulk details of the space-time. The Kerr metric would lead to an axial symmetric angular part (rather than a spherically symmetric one), while the black hole charge gives a sub-leading contribution with respect to, for instance, the mass.
We have hence described the formal steps towards the derivation of the vector field $X^a$. However, upon relying on its defining axioms, it turns out that the simplest candidate is given by 
\begin{equation}
    X=\partial_u,
    \label{X simplest ansatz}
\end{equation}
since it satisfies all the requirements previously discussed. Moreover, it turns out that a more general ansatz, such as
\begin{equation}
    X=\omega(u,x^A)\partial_u,
    \label{X omega ansatz}
\end{equation}
leads to the same vector field $\tilde n^a$, introduced in Eq. \eqref{equation defining n tilde}. 
Indeed, Axiom (iii) in the definition of the asymptotic conformal Killing vector field constrains $\mathcal{L}_X(\Omega^2g)$ only on $\mathcal{N}$. For every metric of the form given in Eq. \eqref{metric in Bondi gauge}
\begin{equation}
    \mathcal{L}_{\partial_u}(\Omega^2g)\Big|_{\mathcal{I}^+}=0, \quad \tilde \Gamma^a{}_{uu}\Big|_{\mathcal{I}^+}=0.
\end{equation}
Hence $X=\partial_u$ defines an asymptotic conformal Killing vector field at null infinity for any asymptotically flat space-time, and not only in the space-time class identified by Eq. \eqref{conditions in which we solve asymptotic conformal killing equation}.
We are going to prove it explicitly in the Schwarzschild case.

\section{The asymptotic conformal generator in Schwarzschild space-time}
As stated in the previous section, in order to simplify the mathematics, we consider the simplest non-trivial asymptotically flat space-time: the Schwarzschild one.
In this case, Bondi coordinates $(u,r,x^A)$ correspond to retarded Eddington-Finkelstein coordinates. Upon defining the tortoise coordinate 
\begin{equation}
    \frac{dr_*}{dr}=\left(1-\frac{2M}{r}\right)^{-1},
\end{equation}
the retarded time is simply given by
\begin{equation}
    u=t-r_*
    \label{retarded time schwarzschild}.
\end{equation}
The metric is obtained by inserting the constraints
\begin{equation}
    m(u,x^A)=M, \quad  C_{AB}=0, \quad  N_A=0,
\end{equation}
 in Eq. \eqref{metric in Bondi gauge}, leading to
\begin{align}
ds^2=&-\left(1-\frac{2M}{r}\right)du^2-2dudr \nonumber\\&+r^2\gamma_{AB}dx^Adx^B.
\label{Shcwarzschild metric in Bondi gauge}
\end{align} For later convenience, we introduce the compactified Schwarzschild metric (quantities referring to such a metric will be denoted with a tilde)
\begin{equation}
    d\tilde s^2=\Omega^2 ds^2,
\end{equation}
which in the $(u,\Omega, x^A)$ chart reads as
\begin{equation}
    d\tilde s^2=-\Omega^2(1-2M\Omega)du^2+2dud\Omega +\gamma_{AB}dx^Adx^B,
    \label{schwarzschil metric in regular chart at infinity}
\end{equation}
where the canonical choice for the conformal factor is $\Omega=1/r$. 
Such a chart is regular at $\mathcal{I}^+$, identified by the condition $\Omega=0$. For later convenience, we highlight that the only non-vanishing Christoffel symbol with upper index $u$ is
\begin{equation}
    \tilde\Gamma^u{}_{uu}=\Omega(1-3M\Omega),
    \label{tilde christtofel uuu}
\end{equation}
while 
\begin{align}
    \tilde \Gamma^\Omega{}_{uu}=&\Omega^3(1-2M\Omega)(1-3M\Omega),\label{Omega christoffel}\\
    \tilde \Gamma^A{}_{uu}=&0.\label{A christoffel}
\end{align}
The components of the induced metric at null infinity are simply  $\tilde h_{AB}=\gamma _{AB}$.
Eqs \eqref{first PDE for X}-\eqref{third PDE for X}  in Schwarzschild space-time read as 
\begin{align}
    2\partial_uX^u+2\left(1+\frac{2M}{r}\right)\partial_uX^r +\frac{2M}{r^2}X^r&=\Lambda+O(r^{-2}),\label{first PDE for X schwarzschild}\\
    \partial_uX^u+\partial_rX^r=&\Lambda,\label{second PDE for X schwarzschild}\\
(\mathcal{L}_X\gamma)_{AB}=&\left(\Lambda-\frac{2}{r}X^r\right)\gamma_{AB}.\label{third PDE for X schwarzschild}
\end{align}
The simplest ansatz given in Eq. \eqref{X simplest ansatz} solves such equations. 
Upon considering $X^a$ given in Eq. \eqref{X omega ansatz}, we obtain the constraint $\omega(u,x^A)=\omega(x^A)$ plus a non-vanishing angular component $X^A$. Indeed, this last solution can be derived by directly integrating the defining PDEs of $X^a$. After a straightforward computation we obtain the leading-order solution
\begin{align}
    X=&\frac{1}{2}\nabla_AY^A(u\partial_u+r\partial_r)\nonumber \\&+\omega(x^A)\partial_u+Y^A\partial_A,
\end{align}
where $Y^A$ is a conformal Killing vector field on the two-sphere. 
Furthermore, we have to impose that its norm vanishes at $\mathcal{I}^+$, i.e.,
\begin{equation}
    \lim_{r\rightarrow\infty}\tilde g_{ab}X^aX^b=0.
\end{equation}
This condition leads to
\begin{equation}
    \gamma_{AB}Y^AY^B=0\implies Y^A=0,
\end{equation}
(since $\gamma_{AB}$ is not degenerate), obtaining simply
\begin{equation}
    X=\omega(x^A)\partial_u-\frac{1}{r}D^A\omega \partial_A,
\end{equation}
where $\omega(x^A)\in C^\infty(S^2)$ (required by the condition of complete integral lines). Since $X^a$ is tangent to the generators of $\mathcal{I}^+$, the affinely parametrised generator $\tilde n^a$ in Eq. \eqref{equation defining n tilde} is fixed on $\mathcal{I}^+$ by the following lemma.\\
\textbf{Lemma}. Let $X^a\big|_{\mathcal{I}^+}=F(u,x^A)\partial_u$. Then, on $\mathcal{I}^+$,
\begin{equation}
    X^b\tilde \nabla_bX^a=(\partial_u F)X^a,
\end{equation}
i.e., 
\begin{equation}
    k\big|_{\mathcal{I^+}}=\partial_uF.
\end{equation}
\textit{Proof}. In the chart of Eq. \eqref{schwarzschil metric in regular chart at infinity}, on $\mathcal{I^+}$, we have that
\begin{equation}
    X^\Omega=X^A=0 ,
\end{equation}
then 
\begin{equation}
    X^b\tilde \nabla_b X^a\Big|_{\mathcal{I}^+}=F\partial_uX^a+F^2\tilde \Gamma^a{}_{uu}.
\end{equation}
Upon considering Eqs \eqref{tilde christtofel uuu}-\eqref{A christoffel}, we conclude that $\tilde \Gamma^a{}_{uu}\Big|_{\mathcal{I}^+}=0$ for every $a$. $\Box$\\
Solving $\mathcal{L}_X\sigma\Big|_{\mathcal{I}^+}=k$ gives
\begin{equation}
    \sigma=\ln|F|+G(x^A), \quad G(x^A)\in C^\infty(S^2),
\end{equation}
and Eq. \eqref{equation defining n tilde} yields
\begin{equation}
    \tilde n\Big|_{\mathcal{I}^+}=e^{-G(x^A)}\partial_u
    \label{general tilde n at scri plus}
\end{equation}
for every $X^a$ tangent to the generators, in particular for $X=\partial_u$ (i.e., $F=1$) and $X=\omega(x^A)\partial_u$. The $G(x^A)$ function encodes the freedom described by the equivalence relation in Eq. \eqref{equivalence class of Carrollian structures}. Its infinitesimal form, given by Eqs \eqref{lee derivative of the metric at null infinity wrt xi}, \eqref{lee derivative of tilde n wrt xi definition}, allows us to compute the generator of asymptotic conformal symmetries. From Eq. \eqref{lee derivative of the metric at null infinity wrt xi} we obtain
\begin{equation}
    \xi^A=Y^A(x^B),
\end{equation}
where $Y^A$ is the conformal Killing vector field of $S^2$ and, moreover, $\mu=D_AY^A$. The other components are obtained from Eqs \eqref{lee derivative of tilde n wrt xi definition}, \eqref{general tilde n at scri plus}. Indeed,
\begin{equation}
    [\xi, \tilde n]=-(Y^A\partial_AG+\partial_u\xi^u)\tilde n-e^{-G}\partial_u\xi^A\partial_A,
\end{equation}
then $\lambda=-Y^A\partial_AG-\partial_u\xi^u$. As already pointed out (cf. Eq. \eqref{rescaling tilde n}), we must have that $\lambda\in C^\infty(S^2)$, implying that $\partial^2_u\xi^u=0$. The final result is then
\begin{equation}
\xi=[f(x^A)u+g(x^A)]\partial_u+Y^A(x^B)\partial_A,
    \label{final xi vector}
\end{equation}
with
\begin{align}
    \mu=&D_AY^A, \label{mu rescaling angular metric}\\
    \lambda=&-f-Y^A\partial_AG.
\end{align}
The two functions $f,g \in C^\infty(S^2)$ are arbitrary. Eq. \eqref{equivalence class of Carrollian structures} allows us to set $G=0$, and then $\lambda=-f$.

\paragraph{Bulk extension and action on the metric} We stress that $\xi^a$  is defined only at $\mathcal{I}^+$. Moreover, its bulk extension is not unique.  
Indeed, the first two Bondi gauge conditions in Eqs \eqref{Bondi gauge conditions} fix $\xi^u, \xi^A$, while the radial component exhibits a free coefficient $\rho \in C^\infty(S^2)$ such that
\begin{align}
    \xi=&(fu+g)\partial_u+\left(Y^A-\frac{1}{r}D^A\xi^u\right)\partial_A\nonumber\\&+\left(\rho r+\frac{1}{2}D_AD^A\xi^u+O(1/r)\right)\partial_r. \label{xi bulk extension}
\end{align}
Since $\xi^\Omega=-\Omega^2\xi^r$, every $\rho$ gives a field tangent to $\mathcal{I}^+$. In the Schwarzschild geometry of Eq. \eqref{Shcwarzschild metric in Bondi gauge}, one finds up to $O(r^{-2})$
\begin{align}
    (\mathcal{L}_\xi g)_{ur}=&-(f+\rho),\\
    g^{AB}(\mathcal{L}_\xi g)_{AB}=&4\rho+2D_CY^C,
\end{align}
therefore the angular boundary metric $q_{AB}$ changes by
\begin{equation}
    \delta q_{AB}=(2\rho+D_CY^C)\gamma_{AB}.
\end{equation}
The two Bondi gauge conditions, $g_{ur}=-1,\delta q_{AB}=0$, hold together only if $f=\frac{1}{2}D_CY^C$, i.e., only for BMS generators. Superdilations necessarily relax one of the Bondi boundary conditions\footnote{On studies of the partial Bondi gauge, cf., e.g., Ref \cite{Geiller:2022vto}}. In the regular chart $(u,\Omega,x^A)$, the effect of superdilations on the compactified metric of Eq. \eqref{schwarzschil metric in regular chart at infinity} at null infinity is given by
\begin{align}
    (\mathcal{L}_\xi \tilde g)_{u\Omega}\Big|_{\mathcal{I}^+}=&f-\rho,\label{first compactified variation}\\
        (\mathcal{L}_\xi \tilde g)_{AB}\Big|_{\mathcal{I}^+}=&D_CY^C \gamma_{AB}. \label{second compactified variation}
\end{align}
For $\rho=f$, only the angular part survives, in agreement with the $\mu$ rescaling of Eq. \eqref{mu rescaling angular metric}. In the $\rho=-f$ case, instead, the condition $g_{ur}=-1$ is preserved and superdilations act as a Weyl rescaling of the boundary metric:
\begin{equation}
     \delta q_{AB}=2w q_{AB},
\end{equation}
with $w$ given by Eq. \eqref{w coefficient}. This is the setting of the Weyl BMS group (cf. Ref. \cite{Freidel:2021fxf}). Finally, Eqs \eqref{first compactified variation}, \eqref{second compactified variation} tell us that $\xi^a$ is a conformal Killing vector of $\tilde g$ on $\mathcal{I}^+$ if and only if $f-\rho=D_AY^A$. For $\rho=-f$, this is nothing but the BMS condition $f=\frac{1}{2}D_AY^A$. Superdilations are conformal symmetries of the Carrollian data $(\tilde h, \tilde n)$ but not of $\tilde g\big|_{\mathcal{I}^+}$, which is why they escape the usual characterisation of BMS.

\subsection{Asymptotic generators algebra}

By denoting with the sub-indices $T,R,D$ the super - translations, rotations and dilations sector, respectively, we can consider the full vector field $\xi^a$ in Eq. \eqref{final xi vector} as the sum of the infinitesimal generators
\begin{align}
    \xi_T(g)=&g\partial_u,\\
    \xi_R(Y)=&Y^A\partial_A,\\
    \xi_D( f)=&fu\partial_u.
\end{align}
The generators $\xi_T(g)$ and $\xi_R(Y)$ close the algebra
\begin{align}
    [\xi_T(g_1), \xi_T(g_2)] &= 0,  \label{commutator two supertrnalsations}\\
    [\xi_R(Y_1), \xi_R(Y_2)] &= \xi_R([Y_1, Y_2]), \\
    [\xi_R(Y), \xi_T(g)]   &= \xi_T(Y[g]).
\end{align}
Supertranslations therefore form an Abelian ideal. The superdilations sector reads
\begin{align}
    \left[\xi_D(f_1),\xi_D(f_2)\right]=&0,\\
    \left[\xi_T(g),\xi_D(f)\right]=&\xi_T(gf),\\
    \left[\xi_R(Y),\xi_D(f)\right]=&\xi_D(Y[f]),\label{commutator R D}
\end{align}
forming an Abelian subalgebra but not an ideal.
Another ideal is given by $T\oplus D$.
We highlight that the split into $T$ and $D$ depends on the origin of $u$, just as the split of Poincar\'e into Lorentz transformations and translations depends on the origin. Indeed, a superdilation about the $u=u_0$ cut can be written as
\begin{align}
    \xi^{u_0}_D (f)=&f(u-u_0)\partial_u \nonumber\\
    =& \xi_D(f)-\xi_T(u_0 f). \label{superdilation generator relative to an origin}
\end{align}
\paragraph{The group structure} At finite level, each generator undergoes an affine map given by
\begin{equation}
    u\rightarrow e^{a(x^A)}u+b(x^A), \quad a,b\in C^\infty(S^2),
\end{equation}
with composition law
\begin{equation}
    (a_1,b_1)\cdot(a_2,b_2)=(a_1+a_2, e^{a_1}b_2+b_1).
\end{equation}
Such a map, together with the conformal maps of the sphere gives
\begin{align}
\Xi_{\mathcal{I}^+}=&\text{Conf}(S^2) \ltimes C^\infty(S^2,\text{Aff}^+(1)) \nonumber\\\cong &\text{Conf}(S^2)\ltimes
\big( C^\infty(S^2) \ltimes C^\infty(S^2) \big).
\label{asymptotic conformal group Scwharzschild}
\end{align}
which is a realisation of the group structure described in Eq. \eqref{symmetry group for ACKH} . We thus recover in asymptotically flat space-times the same algebra derived for superdilations in conformally flat space-times (through a coordinates-based approach) by Donnay et al. (cf. Eqs (2.24) in Ref.  \cite{Donnay:2020fof}).

\paragraph{A comparison with BMS transformations}

Having obtained our main result, given by Eq. \eqref{final xi vector}, we can better understand the core difference between the canonical compactification described in Sec. II and the geometrical construction of asymptotic conformal Killing horizons of Sec. IV. Upon considering a change in the conformal factor 
\begin{equation}
    \Omega\rightarrow\omega \Omega, \quad \omega\in C^\infty(S^2),
\end{equation}
one has that the induced metric at $\mathcal{I}^+$ and the canonical counterpart of our vector field $\tilde n^a$ at $\mathcal{I}^+$ changes as

\begin{align}
    \tilde h \rightarrow& \omega^2 \tilde h\\
    \tilde n^a=\tilde g^{ab}\nabla_b \Omega\rightarrow &\omega^{-1}\tilde n,
\end{align}
fixing $\sigma =\omega^{-1}$, as already mentioned after Eq. \eqref{equivalence class of Carrollian structures}. Infinitesimally, this corresponds to 
\begin{equation}
    \lambda =-\frac{\mu}{2},
\end{equation}
leading to
\begin{equation}
    \xi^u=g+\frac{u}{2}D_AY^A,
\end{equation}
which is exactly $\xi^u_{\text{BMS}}$ at $\mathcal{I}^+$.
The BMS generators are thus embedded as 
\begin{equation}
    \xi_{\text{BMS}}(g,Y)=\xi_T(g)+\xi_R(Y)+\xi_D\left(\frac{1}{2}D_AY^A\right),
\end{equation}
and Eqs \eqref{commutator two supertrnalsations}-\eqref{commutator R D} give
\begin{align}
  &  \left[\xi_R(Y)+\xi_D\left(\frac{1}{2}D_AY^A\right),\xi_T(g)\right]\nonumber\\&=\xi_T\left(Y^A\partial_Ag-\frac{1}{2}gD_CY^C\right)\\\
     &\left[\xi_R(Y_1)+\xi_D\left(\frac{1}{2}D_AY^A_1\right),\xi_R(Y_2)+\xi_D\left(\frac{1}{2}D_AY^A_2\right)\right]\nonumber\\&=\xi_R([Y_1,Y_2])+\xi_D\left(\frac{1}{2}D_A[Y_1,Y_2]^A\right).
\end{align}
We have therefore explicitly shown how superdilations are precisely the relaxation of the canonical relation $\sigma=\omega^{-1}$, the founding principle of asymptotic conformal Killing horizons. 
Superdilations modulo BMS transformations are parametrised by
\begin{equation}
    w\equiv \frac{1}{2}D_AY^A-f \in C^\infty(S^2),
    \label{w coefficient}
\end{equation}
i.e.,
\begin{equation}
    \xi^a=\xi_{\text{BMS}}^a(g,Y)-\xi^a_D(w).
\end{equation}
Notably, BMS transformations act on $w$ as a scalar, i.e.,
\begin{equation}
    [\xi_{\text{BMS}}(g,Y),\xi_D(w)]=\xi_D(Y(w))+\xi_T(wg).
\end{equation}

\paragraph{The Weyl scalar}
Besides the relation with the BMS symmetry group, a notable perspective emerges from the spinorial description of space-times (cf. Refs \cite{Penrose:1960eq, Newman:1961qr, Penrose_Rindler_1984}).  The only non-zero component of the Weyl tensor in Schwarzschild space-time (and in other stationary black hole solutions such as the Kerr-Newman one) is 
\begin{equation}
    \Psi_2=-\frac{M}{r^3},
\end{equation}
which describes that Schwarzschild space-time is of algebraic type D, and that it becomes flat (algebraic type O) as $r\rightarrow \infty$. This transition between algebraic types in the asymptotic region is related to the emergence of superdilations as conformal symmetries, allowed only in type O space-times.

\section{Superdilations charge}

In the previous sections, through an explicit computation in Schwarzschild space-time, we understood that the infinitesimal generator of superdilations at $\mathcal{I}^+$  is given by (cf. Eq. \eqref{final xi vector})
\begin{equation}
    \xi_D=f(x^A)u\partial_u.
    \label{xi d in the section superdilations}
\end{equation}
As we already discussed, it is equal to the one found in the literature for conformally-flat space-times (cf. Ref. \cite{Donnay:2020fof}). It is highly non-trivial that the same structure holds also for asymptotically flat space-times, even if theory suggested such a possibility (cf. Ref. \cite{Barnich:2010eb, Barnich:2010eb}). 
Since the derivation of Sec. V relies only on the Carrollian data on $\mathcal{I}^+$, which are the same for every asymptotically flat metric in Bondi gauge, Eq. \eqref{final xi vector} holds beyond Schwarzschild. However, the bulk extension in Eq. \eqref{xi bulk extension} is not universal, and neither is the phase space on which $\xi_D$ acts. As discussed in Sec. V, $\xi_D$ cannot preserve both $g_{ur}=-1$ and the round boundary metric $q_{AB}$: $\rho=-f$ leads to a phase space with a dynamical conformal factor of the boundary metric, while $\rho=0$ yields the $g_{ur}=-1$ conditions relaxed at order $O(1)$. 
The results we consider below do not depend on this choice. 
The subleading terms are fixed by Eqs \eqref{Bondi gauge conditions} once $\rho$ is chosen, and they enter the charge in a way which we will describe below.
In order to prove that superdilations are non-trivial  gauge symmetries, we have to consider the related Iyer-Wald charge charge, relying on the covariant phase space formalism\footnote{The mathematical framework is described in the references provided. For a broader overview of the covariant formulation of classical field theories with worked examples, cf. Ref. \cite{Gieres:2021ekc}.} (cf. Refs \cite{Crnkovic:1987tz,Zuckerman:1986,Ashtekar:1990gc,Lee:1990nz,wald2000general,Julia:2002df,Harlow:2019yfa, Iyer:1994ys, wald2000general}) . This is necessary since $\xi_D$ is not a proper Killing vector field.
We consider a generic asymptotically flat space-time in General Relativity, following the conventions of Ref. \cite{Compere:2019qed} ($G$ is restored),  The Einstein-Hilbert Lagrangian scalar density in General Relativity is given by
\begin{equation}
    L=\frac{1}{16\pi G}\sqrt{-g}R.
\end{equation}
Under the action of a local variation of the metric (with $\delta$ Grassmann even)
\begin{equation}
    h_{ab}=\delta g_{ab}
\end{equation}
we have
\begin{equation}
    \delta L=-\frac{\sqrt{-g}}{16\pi G}G^{ab}h_{ab}+\partial_a\Theta^a[h;g],
\end{equation}
where the last term is called pre-symplectic potential and reads as
\begin{align}
    \Theta^a[h;g]=&\frac{\sqrt{-g}}{16\pi G}(\nabla_bh^{ab}-\nabla^ah^b{}_b)\nonumber\\=&\frac{1}{16\pi G}\partial_b(\sqrt{-g}h^{ba})+\sqrt{-g}(\Gamma^a{}_{bc}h^{bc}-g^{ab}\partial_bh^c{}_c).
\end{align}
Besides this key ingredient, in order to compute the charge we need to consider the Komar term
\begin{align}
    Q_\xi^{ab}=&\frac{\sqrt{-g}}{16\pi G}(\nabla^a \xi^b -\nabla^b\xi^a)\nonumber\\=&\frac{\sqrt{-g}}{16\pi G}g^{ac}g^{bd}(\partial_c\xi_d-\partial_d\xi_c).
\end{align}
The Iyer-Wald  two-form is then given by
\begin{equation}
    k^{ab}[h,g]=-\delta Q^{ab}+\xi^a\Theta^b-\xi^b\Theta^a.
    \label{definition of kab}
\end{equation}
The final charge is given by integrating such a quantity. Since we are interested at null infinity, the integration is carried on the asymptotic two-sphere\footnote{For a more generic discussion on symmetries at null boundaries and their charges, cf. Refs \cite{Adami:2021nnf, Harlow:2019yfa}.}, leading to (we exploit the notation $\cancel \delta$, introduced in Ref. \cite{Alessio:2019cae}, to stress that the differential is not exact)
\begin{equation}
    \cancel \delta H_\xi=\int_Sd\theta d\phi k^{ur},
\end{equation}
since the $ur$-component is the only one which geometrically matters. After some tensor algebra, for asymptotically flat space-times in General Relativity, we obtain that the superdilations charge split into an integrable part, denoted by $\mathcal{Q}_D$, and a non-integrable one labelled as $\Xi_D$,
\begin{equation}
    \cancel \delta H_D=\delta \mathcal{Q}_D+\Xi_D,
    \label{superdilation charge}
\end{equation}
where\footnote{$d\Omega=\sin\theta d\theta d\phi$  is the area element of the unit two-sphere. No confusion should arise with the conformal factor or the compactified radial coordinate of previous sections.}
\begin{align}
    \mathcal{Q}_D=&\frac{u}{4\pi G}\int_Sd\Omega fm+\mathcal{R}_D,\label{Q D}\\
    \Xi_D=&\frac{u}{32\pi G}\int_Sd\Omega f N_{AB}\delta C^{AB}+\dots \label{Xi D}
\end{align}
The bulk extension in Eq. \eqref{xi bulk extension} enters only through $\mathcal{R}_D$ and the dots in Eq. \eqref{Xi D} depend on the choice of $\rho$ (i.e., the phase space).
We remark that the decomposition is not uniquely defined, since $N_{AB}\delta C^{AB}=\delta (N_{AB}C^{AB})-C^{AB}\delta N_{AB}$ and one can absorb the $N_{AB}C^{AB}$ term into the definition of $\mathcal{Q}_D$. We have chosen to consider the non-integrable part depending on the news tensor, which is the Wald-Zoupas convention.
Notice that, since the bulk extension of $\xi^a$ depends only through its angular derivatives, apart from $\rho$, and Eq. \eqref{definition of kab} is linear in $\xi^a$ and its first derivatives, the superdilations charge differs from the supertranslations one with parameter $uf$ only through $\mathcal{R}_D$ and the dots of Eq. \eqref{Xi D}.   
To linear order around Minkowski, $\cancel \delta \mathcal{R}_D=0$. In Schwarzschild space-time, $\mathcal{R}_D=0$. Its background-dependent part depends on the phase space chosen and is left for future investigations.

\paragraph{The divergence of the charge}
In order to better understand the $u$-dependence of the superdilations charge, we restrict our analysis to Christodoulou-Klainerman space-times (cf. Ref. \cite{christodoulou1993global}), where the Bondi news tensor at $|u|\rightarrow \infty$ falls off as
\begin{equation}
    N_{AB}=O\left(|u|^{-(1+\varepsilon)}\right), \quad \varepsilon>0.
\end{equation}
In such a scenario, at the edges of $\mathcal{I}^+$, the integrable part of Eq. \eqref{superdilation charge} diverges with a rate given by
\begin{align}
    4\pi G\partial_u(\mathcal{Q}_D-\mathcal{R}_D)=& \int_Sd\Omega f\left[m\right.\nonumber \\&\left. +u\left(\frac{1}{4}D_AD_BN^{AB}-\frac{1}{8}N_{AB}N^{AB}\right)\right],
\end{align}
and leading to (assuming $\mathcal{R}_D$ is a local functional of $C_{AB}$ and $N_{AB}$ with no explicit $u$-dependence)
\begin{align}
   \lim_{|u|\rightarrow \infty}\partial_u\mathcal{Q}_D\nonumber=\frac{1}{4\pi G}\int_Sd\Omega f   m_\pm,
\end{align}
where $m_\pm$ is the value of the Bondi mass aspect at $u\rightarrow\pm\infty$.
For space-times which start and end in stationary configurations, $m_\pm$ encodes the Bondi mass aspect of the initial and final states. It is generally non-zero, so the linear growth of $\mathcal{Q}_D$ is generic. It is absent only if $\int_S f m_\pm=0$, e.g., for superdilations without an $\ell=0$ component when the end states are black holes at rest. Remarkably, a divergent superdilations charge is also obtained by Donnay et al. (cf. Sec. 5 of Ref. \cite{Donnay:2020fof}). 
In particular, in studying BMS transformations through an AdS/CFT perspective, some of the bulk Noether charges 
they compute for the bulk superdilations vector field diverge. This disagreement in the bulk-boundary correspondence is seen as an evidence that superdilations are not well behaved asymptotic symmetries. Their analysis regards, as already written, conformally flat space-times. However, there are several connections with ours which leads to a wider description of superdilations throughout different kind of space-times.

\paragraph{Action on radiative data} Since superdilations are non-trivial diffeomorphisms but change the Bondi boundary conditions, their action on radiative data is that of a field-dependent change of frame. In Minkowski space-time (i.e., $m=0=C_{AB}$) we find, for every $\rho$,
\begin{align}
    (\mathcal{L}_D C)_{AB}=&-u(2D_AD_B-\gamma_{AB}D_CD^C)f,\\
    (\mathcal{L}_DN)_{AB}=&\partial_u(\mathcal{L}_D C)_{AB}.
\end{align}

For $f$ with $\ell\ge2$ components, we have a pure-gauge but non-vanishing Bondi news tensor, analogous to the one generated by superboosts (cf. Ref. \cite{Compere:2018ylh}). Superdilations therefore do not preserve the vacuum $N_{AB}=0$ sector, and including them requires an extended phase space. In the Schwarzschild geometry, the Bondi mass aspect is changed by
\begin{equation}
    \delta_{\xi_D}m=(2-\kappa)fM,
\end{equation}
for $\rho=\kappa f$. For the global dilation mode ($\ell=0, \kappa=1$), we have $\delta_{\xi_D}m=fM$,

\subsection{The charge in Schwarzschild space-time} While postponing a more complete discussion to a future work, we can study the charge in Schwarzschild space-time, where it is integrable and reads as
\begin{align}
    H_D^{\text{Sch}}(u,f)=&\frac{M}{4\pi G}u\int_Sd\Omega \,f\nonumber\\=& \frac{f_{00}}{\sqrt{4\pi}}\frac{M}{G}u.\label{superdilations charge Schwarzschild}
\end{align}
In the last equality we considered that, since $f\in C^\infty(S^2)$, it can be expanded with respect to spherical harmonics
\begin{equation}
    f(x^A)=\sum_{\ell=0}^\infty\sum_{m=-\ell}^\ell f_{\ell m}Y_\ell^m(x^A), \quad f_{\ell m}\in \mathbb{R},
\end{equation}
leading to a trivial integration over the two-sphere:
\begin{equation}
    \int_Sd\Omega f=\sqrt{4\pi}f_{00}.
\end{equation}
Eq. \eqref{superdilations charge Schwarzschild} tells us that, in Schwarzschild  space-time, only the $\ell=0$ mode contributes to the superdilations charge.
The same holds for supertranslations (only the energy is non-vanishing). Indeed, on a fixed cut of the celestial sphere the two charges are equivalent, i.e.,
\begin{equation}
    H_D^{\text{Sch}}(u_0,f)=u_0H_T^{\text{Sch}}(f),
\end{equation}
where
\begin{equation}
    H_T^{\text{Sch}}(g)=\frac{g_{00}}{\sqrt{4\pi} }\frac{M}{G}.
\end{equation}
The charge associated to a superdilation with respect to a fixed cut of $\mathcal{I}^+$ (cf. Eq. \eqref{superdilation generator relative to an origin}), is instead given by
\begin{equation}
    H_D^{\text{Sch}}(u-u_0,f)=H_D^{\text{Sch}}(u,f)-u_0H_T^{\text{Sch}}(f).
\end{equation}
Since in Schwarzschild space-time $C_{AB}=0$ and $\partial_um=0$, no energy is radiated through $\mathcal{I}^+$. The $u$-divergence of the charge therefore does not depend on energy fluxes but it is intrinsic in its definition. It stems from $\partial_u\xi^u_D=f$, in the same way as the explicit time dependence of boost or dilation charges in Minkowski space-time. Only the $\ell=0$ mode, i.e., a global dilation, contributes and it is not a proper bulk isometry of Schwarzschild. Such a divergence forbids the antipodal matching at $i^0$, since, upon considering the superdilations charge at past null infinity
\begin{equation}
    H_D^{(-)\,\text{Sch}}(v,f^{(-)})=\frac{f^{(-)}_{00}}{\sqrt{4\pi}}\frac{M}{G}v,
\end{equation}
it clearly diverges with opposite sign with respect to $H_D^\text{Sch}$. 
On $t=const$, $u=t-r_*$ and $v=t+r_*$, hence the divergent pieces at $\mathcal{I}^+_-$ and $\mathcal{I}^-_+$ have opposite signs and depend only on the direction of approach to $i^0$.
Since $H^{\rm Sch}_D$ is exactly linear in $u$, the subtraction
$H^{\rm Sch}_D-u\,\partial_uH^{\rm Sch}_D$ vanishes identically, and the only
$u$-independent quantity carried by a superdilation is its rate,
\begin{equation}
\partial_u H^{\rm Sch}_D(u,f)=H^{\rm Sch}_T(f),
\end{equation}
which is finite at both $\mathcal{ I}^+_\pm$ and coincides with the
supertranslation charge with parameter $f$. With the antipodal identification
$f^{(+)}(\hat x)=f^{(-)}(-\hat x)$, the matching of the rates,
\begin{equation}
\partial_u H^{(+)\rm Sch}_D\Big|_{\mathcal {I}^+_-}
=\partial_v H^{(-)\rm Sch}_D\Big|_{\mathcal{ I}^-_+},
\end{equation}
reduces to $f^{(+)}_{00}=f^{(-)}_{00}$, which holds automatically since
$Y_{00}$ is even under the antipodal map. Hence, in the Schwarzschild geometry,
the superdilation charge carries no information beyond the monopole
supertranslation charge.
It is therefore mandatory to investigate radiative phase space to go beyond such issues.

\section{Superdilated BMS observers}

In order to relate superdilations to some kind of observable effects, we introduce a family of BMS observers as in Ref. \cite{Strominger:2014pwa}, i.e., detectors which travel along worldlines at fixed radius (assumed large) and angles. They can be labelled by the vector field
\begin{equation}
    X^a_{\text{BMS}}(s)=(s,r_0,x^A_0),
\end{equation}
where $s$ is the affine parameter on the worldline they describe. At first order in $r$, BMS observers are nearly inertial. They provide a useful tool to investigate physical effects of superdilations at null infinity. Let us consider, for instance, two generic BMS observers located on the same celestial sphere but slightly displaced in retarded time and angle variables. The displacement vector, which we assume with constant components, is given by
\begin{equation}
    \zeta=\delta u\partial_u +\delta x^A\partial_A.
\end{equation}
The action of standard BMS supertranslations on such a vector leads to the displacement memory effect (cf. Eq $(4.11)$ in Ref. \cite{Strominger:2014pwa}). 
The action of  $\xi^a$ (cf. Eq. \eqref{final xi vector})
on these two observers is instead given by the Lie derivative
\begin{align}
  \mathcal{L}_{\xi}\zeta=&[\xi,\zeta] \nonumber\\=&-\left[  z(x^A)\delta u+\mathbf{\Theta}_B(u,x^A)\delta x^B\right]\partial_u \nonumber\\&-\delta x^B\partial_BY^A\partial_A,
\end{align}
where 
\begin{align}
z(x^A)=&f(x^A),\\
       \mathbf{\Theta}_B(u,x^A)=&u \partial_B f(x^A)+\partial_Bg(x^A).
\end{align}
The non-vanishing of the Lie derivative means that, as expected, the displacement is not invariant under the action of $\xi^a$ and, in particular, of superdilations.
In Schwarzschild space-time, upon defining the local charge density as
\begin{equation}
    q=\frac{M}{4\pi G}(uf+g),
\end{equation}
we can understand the two functions $z(x^A)$ and $\mathbf{\Theta}_B$ as local charge-induced effects, since

\begin{align}
z(x^A)=&\frac{4\pi G}{M}\partial_u q,\\
       \mathbf{\Theta}_B(u,x^A)=&\frac{4\pi G}{M}\partial_B q.
\end{align}
Upon considering the superdilations sector, we can recognise two different contributions.\\
Upon setting $\delta u=0$, we focus on two synchronized observers located at two different points of the celestial sphere. The $\mathbf{\Theta}_B(u,x^A)$ factor encodes the signature of superdilations: different observers are effectively pushed by the dilation at different rates, depending on their position. A similar phenomenon happens for supertranslations. \\ The case in which $\delta x^A=0$ corresponds, instead, to a single observer who measures two nearby light pulses with a retarded time difference equals to $\delta u$. We can better understand this phenomenon by considering an observer which measures the arrival time of two light pulses, each of them coming from two different points of null infinity. Since the space-time we are considering is spherically symmetric, it should be equal since such pulses come from the same celestial sphere. However, as the computation shows, the measurement of the arrival time of the signals also depends on their angular position. Since $\mathcal{L}_\xi \zeta$ is a coordinate statement, a genuine memory-type observable requires a physical change between configurations, such as the frame change of Sec. VI. We leave this to future investigations.

\section{Conclusions and outlook}

In this paper, following the theoretical description given in Ref. \cite{deAguiarAlves:2025dce}, we explicitly derive the generator of asymptotic conformal symmetries at null infinity, given in Eq. \eqref{final xi vector}, in Schwarzschild space-time. Besides the supertranslations and Lorentz transformations, the superdilations sector extends the BMS algebra. As discussed in Sec. III, the existence of superdilations in asymptotically flat space-times is highly non-trivial, since these geometries rule out bulk conformal transformations such as dilations. Their emergence as asymptotic symmetries does not have therefore a bulk correspondence. Moreover, they have been studied only in conformally flat space-times or expanding universe models. Even if Schwarzschild is the simplest example of asymptotically flat space-time, we can, nonetheless, observe unexpected features related to superdilations. The most interesting one is the existence of a non-trivial charge, whose linear growth in retarded time reflects the explicit time dependence of the generator rather than a flux, and whose effect on BMS observers is an angle-dependent dilation rate. However, an open problem is the divergence of the charge (even in the most general case) in the $|u|\rightarrow \infty$ limit. Even if we provided a preliminary analysis on such an issue, it is clear that a deeper investigation is required, which goes beyond the scope of this paper. \\A number of potential extensions may therefore be identified to broaden the scope of our studies. First of all, it could be interesting to compute bulk terms of the asymptotic conformal generator $\xi^a$ of Eq. \eqref{final xi vector} for general asymptotically flat space-times, in order to properly compare superdilations with the other BMS symmetries.
The effect of superdilations on gravitational radiation needs more attention.
This is motivated by a direct comparison with supertranslations: their physical consequences are largely encoded in the gravitational memory effect, which is strictly related to the shear tensor and to a well-defined charge. Indeed, the leading soft graviton theorem (related to the memory effect) arises from differences of supertranslations charges before and after scattering. From this infrared-gravity perspective, a divergent charge introduces ambiguity into this relationship. From another perspective, a well-defined charge establishes a connection with black hole thermodynamics. Indeed, the quantity $2\pi H_\xi$ could be understood as an asymptotic conformal black hole entropy by generalising (in an appropriate way) the considerations made by Wald in Refs \cite{Wald:1993nt, Iyer:1994ys}.
\\Provided that all of the above issues are properly addressed, the study of superdilations can follow the avenue paved by other BMS symmetries. One of the most relevant research proposal is the construction of particles states (as done in Ref. \cite{Bekaert:2024uuy} for BMS particles), a quantum Hilbert space for semi-classical investigations (cf. Ref. \cite{Donnay:2020fof} for conformally flat space-times), and soft-hair models for black holes (as pointed out in Ref. \cite{Hawking:2016msc}). Furthermore, it would be interesting to understand the role of superdilations in the infrared gravity research program. While dilations symmetry is related to short-distances effects in standard quantum field theory, in the description of the asymptotic region of Schwarzschild space-time its role is completely different. Superdilations are asymptotic conformal diffeomorphisms  (therefore more general than asymptotic isometries), emerging from a gauge freedom in choosing the conformal scaling among the induced metric at null infinity and the tangent vector field to null generators of null infinity (cf. Eq. \eqref{equivalence class of Carrollian structures}). Such a freedom is absent in the canonical construction of null infinity. The role played by superdilations in the infrared region of gravity relies on this aspect. Indeed, null infinity is, by construction (cf. Ref. \cite{Penrose1963}), defined through conformal methods and hence one should consider also asymptotic conformal diffeomorphisms in order to have a complete description of asymptotic symmetries. We can point out another possible motivation to consider seriously the emergence of superdilations. In conformally flat space-times with a cosmological horizon, asymptotic isometries are relevant in the construction of Hadamard states, as clearly depicted in Ref. \cite{Dappiaggi:2017kka}. An analysis in the conformal case (asymptotically flat space-times) could then be possible starting from the results of this paper. Each one of these topics deserves particular attention.
However, they all pave the way for a completely new research line.

\begin{acknowledgments}
I would like to thank Prof. Michele Arzano for his valuable advice and helpful discussions throughout the preparation of this work.\\
I acknowledge support from the INFN
Iniziativa Specifica QUAGRAP and from the European COST actions BridgeQG CA23130
and CaLISTA CA21109.\\
I dedicate this paper to my brother Sergio.
\end{acknowledgments}

\bibliography{apssamp}

\end{document}